\documentclass[journal,twocolumn]{IEEEtran}
\usepackage{epsfig,makeidx,color}
\usepackage{amsmath,amssymb,bbm}
\usepackage{cite,graphicx,lipsum}
\usepackage{enumerate}
\usepackage[switch,pagewise]{lineno}
\usepackage{hyperref}
\hypersetup{
        colorlinks = true,
        citecolor=blue,
}
\newcommand{\indicator}{\mathbbm{1}}

\def\bpsi{\mbox{\boldmath $\psi$}}

\def\cQ{{\cal Q}}

\def\rH{{\rm H}}
\def\rT{{\rm T}}

\def\uC{{\mathbb C}}
\def\uE{{\mathbb E}}
\def\uW{{\mathbb W}}

\newtheorem{mylemma}{\bf Lemma} 

\def\be{ \begin{equation} }
\def\ee{ \end{equation} }
\def\bea{ \begin{eqnarray} }
\def\eea{ \end{eqnarray} }

\def\bx{{\bf x}}
\def\by{{\bf y}}

\def\bs{{\bf s}}
\def\ba{{\bf a}}
\def\br{{\bf r}}

\def\bn{{\bf n}}

\def\bI{{\bf I}}

\def\bS{{\bf S}}

\def\bR{{\bf R}}

\def\b0{{\bf 0}}

\def\bPsi{{\bf \Psi}}

\def\cC{{\cal C}}

\def\cB{{\cal B}}

\def\cQ{{\cal Q}}

\def\cN{{\cal N}}
\def\cS{{\cal S}}

\ifCLASSOPTIONonecolumn
  \newcommand{\figwidth}{0.55\columnwidth}
  
\else
  \newcommand{\figwidth}{0.85\columnwidth}
  
\fi

\begin{document}

\title{On Throughput of Compressive Random Access for
One Short Message Delivery in IoT}

\author{Jinho Choi\\
\thanks{The author is with
the School of Information Technology,
Deakin University, Geelong, VIC 3220, Australia
(e-mail: jinho.choi@deakin.edu.au).
This research was supported 
by the Australian Government through the Australian Research 
Council's Discovery Projects funding scheme (DP200100391).}}


\maketitle
\begin{abstract}
In this paper, we study compressive
random access (CRA) with two stages for machine-type
communication (MTC) in cellular
Internet-of-Things (IoT).
In particular, we consider the case that each
user (IoT device or sensor)
has only one short message (of the same length) 
when it is activated to send data in IoT applications.
Two different CRA-based random access schemes
are discussed (one is conventional and the other is new
based on a simplified handshaking process). 
Based on the throughput analysis,
we show that the CRA-based random access
scheme with simplified handshaking process
can outperform as its length of payload 
is adaptively decided depending on the number of active users.
Simulation results confirm that
the derived throughput expressions agree with
them and can be used to design a random access system for MTC
with
each active device or sensor that has one short message.
\end{abstract}

\begin{IEEEkeywords}
Machine-Type Communication; 
Compressive Random Access;
the Internet-of-Things
\end{IEEEkeywords}

\ifCLASSOPTIONonecolumn
\baselineskip 28pt
\fi

\section{Introduction}

Recently, the Internet of Things (IoT) has been
extensively studied and 
a number of different applications of the IoT are considered 
for smart cities and factories \cite{Gubbi13} \cite{Kim16}.
To support the connectivity for various IoT applications,
cellular IoT has been considered.
For example, in \cite{Mang16},
a deployment study of narrowband IoT (NB-IoT)
\cite{3GPP_NBIoT} is carried out
to support IoT applications over a large area.
In cellular IoT,
machine-type communication (MTC)
\cite{3GPP_MTC} \cite{3GPP_NBIoT} 
plays a crucial role in providing the connectivity for IoT
devices and sensors \cite{Shar15}.
Due to sparse activity of those devices and sensors,
MTC is usually based on random access 
to keep signaling overhead low 
\cite{3GPP_MTC} \cite{3GPP_NBIoT} 
\cite{Galinina13} \cite{Chang15} \cite{Choi16}.

Due to a large number of IoT devices and sensors,
it is expected to support 
massive connectivity in MTC 
\cite{Bockelmann16}.
To this end, the notion of massive multiple input multiple output (MIMO)
\cite{Marzetta10} can be considered.
In \cite{deC17}  \cite{Senel17} \cite{Liu18} \cite{Liu18a},
massive MIMO based random access schemes
are studied for massive MTC,
where a base station (BS) is equipped with a large number of antenna
elements. 
In \cite{Jiang15} \cite{Ciuonzo15}, 
massive MIMO is used to collect data from sensors
in wireless sensor networks (WSNs).
Based on
\cite{Bjornson18}, it seems that massive MIMO is a solution to 
massive MTC, as the capacity becomes unbounded in the presence
of pilot contamination, which may allow to support a very large 
number of devices in each cell.

However, if a large number of antenna elements
at a BS are not available (due to various reasons including
cost), it may be necessary for BS 
with single antenna to consider other approaches that
have a high spectral efficiency or throughput to support MTC.
In order to improve the throughput of random access for MTC, 
the notion of compressive sensing
has been applied to multiuser detection (MUD)
at a BS (equipped with single antenna)
in \cite{Zhu11} \cite{Applebaum12}.
In \cite{Schepker13}
\cite{Choi17IoT} \cite{Abebe17},
in order to estimate the channel state information (CSI)
for coherent detection, two-stage approaches
are considered for random access,
where in general the first stage is to transmit preambles
or pilots (so that the BS can estimate the CSI of active users)
and the following second stage is to transmit data packets.
The resulting approach is called compressive
random access (CRA) \cite{Wunder14},
as the notion of compressive sensing 
\cite{Donoho06} \cite{Candes06} can be employed
to exploit the sparse activity of users in MUD.
In general, CRA can be seen as a
multichannel random access scheme,
where each (multiple access) channel can be characterized by
a preamble in the first stage.
In the second stage, the preamble can be used
as a spreading sequence for MUD, which results in
code division multiple access (CDMA) based random
access in  \cite{Zhu11} \cite{Applebaum12}.
As shown in \cite{Abebe17}, 
it is possible to use different spreading code (but associated
with the preamble in the first stage) for spreading
in the second stage.

The notion of CRA is applied to an existing random
access scheme for MTC in \cite{Seo19}.
As stated in \cite{Seo19} \cite{Choi_CRA18},
a salient feature of CRA is that the preambles
can be non-orthogonal. If orthogonal preambles are used,
the number of preambles becomes the length of preambles.
However, if non-orthogonal preambles are used,
the number of preambles can be much larger than
the length of preambles, which can effectively reduce
the probability of
preamble collision (PC) and improve the performance
of random access.

In this paper, we study two two-stage random
access schemes based on CRA,
where the first stage is used to transmit preambles by
active users so that the BS is able to estimate CSI
of active users,
while the second stage is used to transmit data packets.
Throughout the paper, a special case is 
considered,
where each active user has only one data packet
of the same length.
This might be the case where each active
user has a short message
(e.g., a few tens bytes) in IoT applications \cite{Naik17},
where a user is an IoT device or sensor.
In general, since the length of data packet
is short, it is desirable to
have a low signaling overhead (which justifies
the use of random access) as well as a high throughput.
We show that CRA-based two-stage random access schemes
can have high throughput (than that of conventional
multichannel ALOHA). 
One of the two CRA-based schemes in this paper is
similar to conventional grant-free CRA-based schemes
in  \cite{Choi17IoT} \cite{Abebe17} (as no access grant
is pursued by users). While
the other scheme differs from the conventional schemes,
it can be seen as a special case of \cite{Seo19} with 
a simplified handshaking process. 
Since
we only consider \emph{one} short packet transmission from
each active user,
a specific implementation of the second stage
becomes possible as will be explained later in this paper.
The resulting scheme can provide a higher throughput
than conventional grant-free CRA-based scheme.
In summary, the main contribution of the paper
is two-fold: \emph{i)} a new CRA-based
random access scheme with 
a simplified handshaking process is proposed,
which might be suitable for IoT applications where
each device or sensor has only one short message to send
in each access;
\emph{ii)} a throughput analysis is studied 
with the length of the second stage
that varies and depends on the number of active users.

Note that the proposed approach in this paper
differs from that in \cite{Abebe17} as the feedback
from the BS is exploited (through a simplified handshaking process) 
and that in \cite{Seo19} 
as the number of slots for data packet transmissions is adaptively decided
to improve the throughput.

The rest of the paper is organized as follows.
In Section~\ref{S:SM}, we present the system model for 
two CRA-based random access schemes.
The throughput of 
two CRA-based random access schemes is analyzed
in Section~\ref{S:Thp}. In Section~\ref{S:OI},
we briefly discuss some other issues that
are not studied in 
Section~\ref{S:Thp}. Simulation
results are presented in Section~\ref{S:Sim},
and the paper is concluded with some remarks in Section~\ref{S:Conc}.

{\it Notation}:
Matrices and vectors are denoted by upper- and lower-case
boldface letters, respectively.
The superscripts $\rT$ and $\rH$
denote the transpose and complex conjugate, respectively.
The support of a vector is denoted by ${\rm supp} (\bx)$
(which is the number of the non-zero elements of $\bx$).
$\uE[\cdot]$
and ${\rm Var}(\cdot)$
denote the statistical expectation and variance, respectively.
$\cC \cN(\ba, \bR)$
represents the distribution of
circularly symmetric complex Gaussian (CSCG)
random vectors with mean vector $\ba$ and
covariance matrix $\bR$.

\section{System Model}	\label{S:SM}

Suppose that a random access system 
consists of a BS and a number of users,
where all the users are synchronized.
In this paper, we consider a particular
case where
each active user has only one packet to send.
For example, in IoT applications with a number of 
environmental sensors,
each sensor may have a fixed small number of bytes
to send to the BS. 
In this case, a complicated approach based
on handshaking (e.g., \cite{3GPP_NBIoT} \cite{3GPP_MTC}) 
might result in unnecessary signaling
overhead.
To avoid it, simple approaches based on
two stages can be considered with 
a pool of pre-determined preambles that are non-orthogonal
sequences. 
In this paper, we consider two slightly different approaches
with two stages.

\subsection{CRA-1}

In this subsection, we consider a 
random access approach with two stages.
Similar to \cite{Abebe17}
in Stage 1, an active user is to randomly choose
a preamble from the pool.
Let $\bpsi_l$ be the $l$th
preamble in the pool of $L$ preambles, $\{\bpsi_1,
\ldots, \bpsi_L\}$, where $L$ represents
the number of available preambles.
Here, $\bpsi_l$ is a sequence of length $N$.
The BS receives the preambles transmitted by
active users and can detect them
using a CS-based approach.

For convenience, denote by $T_{\rm P} = N T_{\rm s}$ the duration
of Stage 1, where $T_{\rm s}$ is the unit symbol duration 
or the inverse of the system bandwidth (with an ideal pulse shaping filter).
For convenience, $T_{\rm s}$ is normalized, i.e.,
$T_{\rm s} = 1$ (as a result, $T_{\rm P} = N$).
We assume that $L \gg N$ (since the preambles
are non-orthogonal, there can be more than $N$ sequences).
The resulting scheme is referred to as 
CRA \cite{Wunder14} in this paper. In CRA,
it is usually assumed that 
the number of active users, denoted by $K$, is sufficiently
small, i.e., $K \ll N$.
In Stage 1, the BS is able to estimate 
the channel coefficients 
of $K$ active users
in conjunction with the detection of transmitted preambles
using CS algorithms.
In this paper, we assume that the bandwidth is sufficiently
narrow so that the channels are modeled as flat fading channels
as in \cite{Applebaum12}
(note that frequency-selective fading channels
are considered in \cite{Schepker13} \cite{Wunder14}).

In Stage 2, each active user transmits its data packet
of $M$-symbol with spreading by the preamble  
that is used in Stage 1 (similar to CDMA as in \cite{Zhu11}
\cite{Applebaum12}).
Thus, the spread packet duration becomes
$N T_{\rm D}$, where  $T_{\rm D} = M T_{\rm s}
= M$ denotes the packet duration (in unit time).
Since the transmitted preambles can be detected in 
Stage 1, they can be used for MUD
\cite{VerduBook}
in Stage 2 to recover data symbols from spread signals.
Note that if $K \le N$, where $N$ is regarded as
the spreading gain, MUD can recover data symbols from
$K$ active users (provided that the signal-to-noise ratio (SNR)
is sufficiently high as well as the correlation between preambles
is sufficiently low).

\begin{figure}[thb]
\begin{center}
\includegraphics[width=\figwidth]{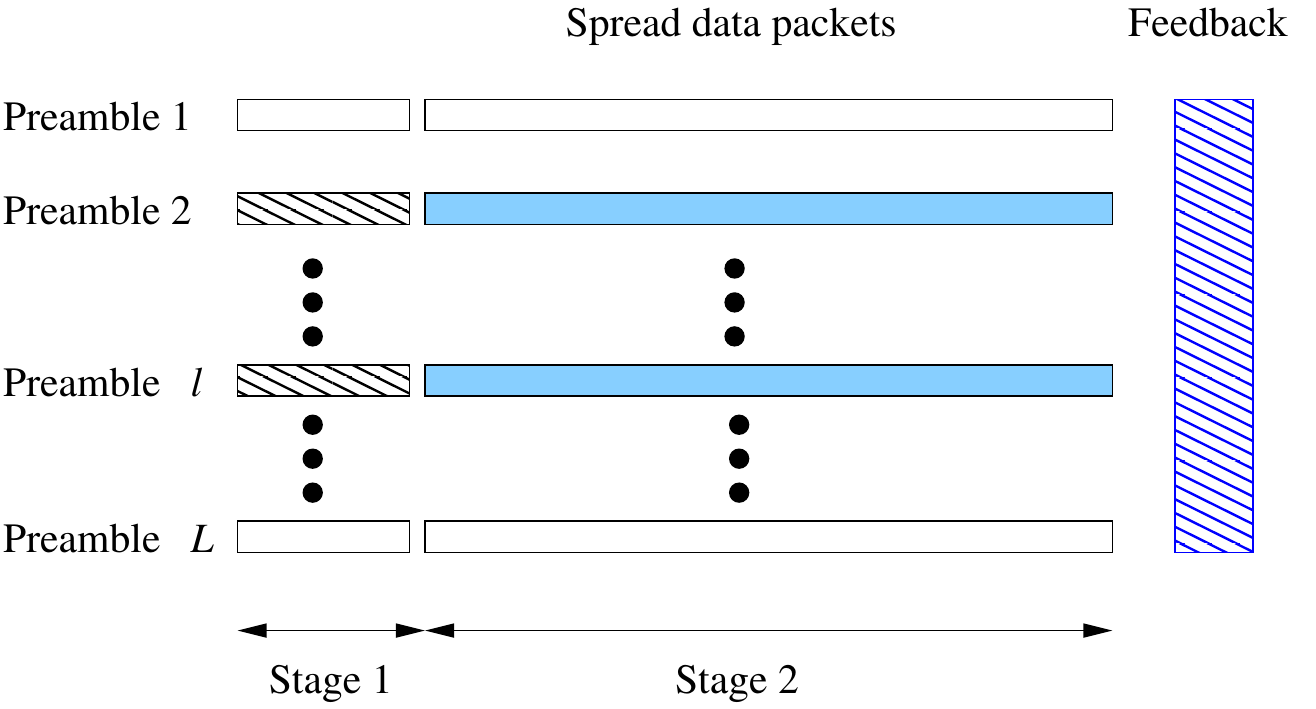}
\end{center}
\caption{CRA-1 with 2 stages, where active users' preambles
are denoted by shaded blocks and 
their data packets are shown in light blue.}
        \label{Fig:cra1}
\end{figure}

For convenience, the resulting approach is
referred to as CRA-1.
In Fig.~\ref{Fig:cra1}, we illustrate a session
consisting of Stage 1 for preamble transmissions
and Stage 2 for data packet transmissions.
Note that at the end of a session, there is a feedback
from the BS to users to inform the success or failure
of data packet decoding.

Let $h_k$ and $l(k)$ denote
the channel coefficient from active user $k$ to the BS
and the preamble index chosen by active user $k$, respectively.
Since there are $K$ active users,
the received signal at the BS during Stage 1 is given by
\begin{align}
\by = \sum_{k=1}^K \bpsi_{l(k)} \sqrt{P_k} h_k + \bn 
= \bPsi \bs + \bn,
	\label{EQ:by}
\end{align}
where $\bPsi = [\bpsi_1 \ \ldots \ \bpsi_L] \in \uC^{N \times L}$,
$\bs$ is a $K$-sparse vector, $P_k$ is the transmit power
of active user $k$,
and $\bn \sim \cC \cN(\b0, N_0 \bI)$ is the background noise vector.
The received signal at the BS during Stage 2 is given by
\begin{align}
\br_m 
& = \sum_{k=1}^K \bpsi_{l(k)} \sqrt{P_k} h_k d_{k,m} + \bn_m \cr
& = \bPsi \bs_m + \bn_m, m = 0, \ldots, M-1,
\end{align}
where $d_{k,m}$ is the $m$th data symbol of data packet from
active user $k$ and 
$\bn_m \sim \cC \cN(\b0, N_0 \bI)$ is the background noise vector.
Here, $\bs_m$ is a $K$-sparse vector that has 
the same support as $\bs$ in \eqref{EQ:by}.
Thus, the received signals during Stage 1 and Stage 2 can be
seen as $[\by^\rT \br_0^\rT \ \ldots \ \br_{M-1}^\rT]^\rT$,
which  is a vector of length $N + M N= (1+M)N$.

In fact, $\{\by, \br_0, \ldots, \br_{M-1}\}$ can be seen
as multiple measurement vectors (MMV)
in the context of compressive sensing
\cite{Chen06} \cite{Davies12}.
It is known that
a sufficient and necessary condition to estimate the support of
$\bs$ or the indices of the transmitted 
preambles
is given by
\be
K < \frac{{\rm spark}(\bPsi) -1 + {\rm rank}(\bS)}{2},
        \label{EQ:Ms}
\ee
where $\bS = [\bs \ \bs_0 \ \ldots \ \bs_{M-1}]$ and
$\operatorname{spark}(\bPsi)$ is
the smallest number of columns from  $\bPsi$ that are linearly dependent
\cite{Donoho03}.
For a random vector for $\bPsi$, 
${\rm spark}(\bPsi) -1 = {\rm rank}(\bPsi) = N$ w.p. 1
\cite{Bruckstein09}. Thus, if ${\rm rank}(\bS)  = N$,
from \eqref{EQ:Ms},
it is possible to detect up to $N-1$ transmitted preambles.
Then, with the detected transmitted preambles,
the channel coefficients can be estimated from $\by$. With
the estimated channel coefficients as well as 
detected transmitted preambles, data packets can be
decoded from $\{\br_0, \ldots, \br_{M-1}\}$.

In \cite{Choi_CRA18},
the performance of CRA is compared with
that of multichannel ALOHA.
In multichannel ALOHA, it is assumed that
the preambles are 
orthogonal and each preamble represents an orthogonal
(multiple access)
channel. As a result, the number of preambles
is equal to the length of preamble, i.e., $L = N$.
In this case, it is not necessary to use CS algorithms, because
each channel is seen as an independent channel and 
a simple correlator becomes optimal to detect the presence
of signal in each channel (i.e., no CS-based MUD is required).
However, 
since the number of preambles in CRA can be higher
than that in multichannel ALOHA, 
it is shown that
the throughout of CRA
can be higher\footnote{This is due to the fact that the
increase of preambles 
reduces the probability of PC.}
than that of multichannel ALOHA by a factor of 2
\cite{Choi_CRA18}.

CRA-1 is often considered to be a grant-free approach
\cite{Abebe17}, because an active user
can transmit a data packet without asking 
a dedicated channel (for payload).

\subsection{CRA-2}

In this subsection, we consider a slightly different approach
from CRA-1, while its Stage 1 is the same as that in CRA-1.

\begin{figure}[thb]
\begin{center}
\includegraphics[width=\figwidth]{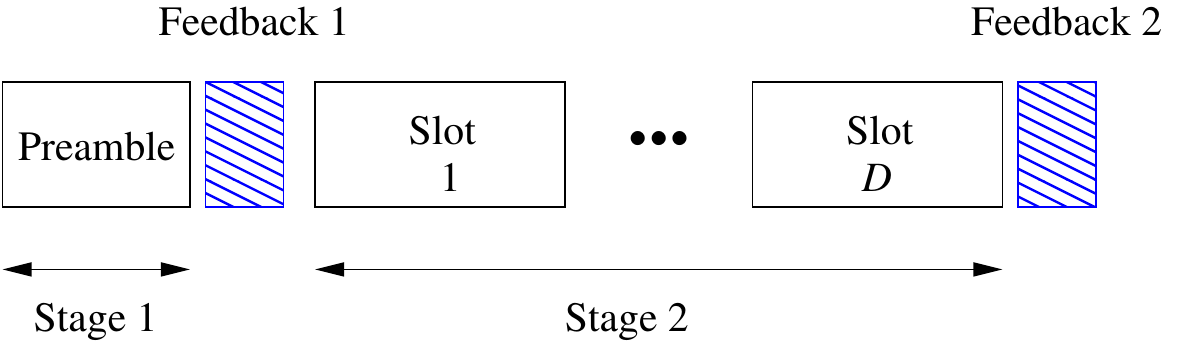}
\end{center}
\caption{CRA-2 with 2 stages (when
each active user has only one data packet to send), where the number of 
slots for uplink data packets from active users, $D$,
varies.}
        \label{Fig:two_stage}
\end{figure}

After Stage 1 
(i.e., once the BS is able to detect the transmitted preambles),
suppose that the BS sends feedback signal to users to inform
the indices of detected transmitted preambles,
which is referred to as Feedback 1.
Let $D$ denote the number of the detected preambles and
denote by $l_d \in 
\{1,\ldots, L\}$ the index of the $d$th detected preamble
with the following increasing order:
$l_1 < l_2 < \ldots < l_D$.
In Stage 2, there are $D$ slots for 
transmissions of data packets. 
The BS expects to receive a data packet during the
$d$th slot from the active user 
transmitting the preamble of index $l_d$.
As a result, the duration of Stage 2 is $D T_{\rm D}
= D M$ as illustrated in Fig.~\ref{Fig:two_stage}.
Note that since $D$ varies,
the length of Stage 2 varies from a session to 
another.
The length of session is the sum
of the lengths of Stages 1 and 2 and Feedbacks 1 and 2.

Since an active user that transmits a preamble, 
say $l_d$, knows that the slot for its data packet is
the $d$th slot from Feedback 1, 
it can transmit a data packet in the $d$th slot in Stage 2.
At the end of Stage 2, the BS sends
another feedback signal to the users to inform 
success or failure of decoding of data packets,
which is referred to as Feedback 2.
The resulting scheme is referred to as CRA-2.

It is noteworthy that CRA-2 can be seen as a
special case of the approach in \cite{Seo19},
where CRA is used only for connection establishment.
Thanks to the presence of Feedback 1, the operation of Stage 1
in CRA-2 is seen as a simplified handshaking process as in 
\cite{Seo19}. Furthermore,
since only one data packet per active user is assumed in this paper,
in Stage 2, channels for payloads
(or data packets) can be easily reserved 
without further requests from active users
(which is not considered in \cite{Seo19}).
Note that if each active user has a different number of
data packets to transmit, 
the length of slot can be set to the maximum length of data packet
in CRA-2 at the cost of degraded spectral efficiency.

For simplicity,
we assume that the duration of Feedback 1 is equal to
that of Feedback 2 and the total duration of feedback 
is denoted by $\tau$.
Note that in CRA-1, there might be feedback
to inform whether or not data packets are successfully decoded.
In this case, CRA-1 has only Feedback 2.
In Table~\ref{TBL:1},
we summarize the 
duration of each stage and feedback in CRA-1 and CRA-2.
Throughout the paper, we assume that $\tau \ll T_{\rm D}$.
A notable difference between CRA-1 and CRA-2 is that
the length of Stage 2 of CRA-2 is varying, but
that of CRA-1 is fixed.
From this, it is expected that CRA-2 can perform
better than CRA-1 when the number of active users, $K$,
is less than the spreading gain
or the length of preambles, $N$.


\begin{table}[thb] 
\caption{The duration of each stage and feedback in CRA-1 and CRA-2.}
\begin{center}
\begin{tabular}{|c||c|c|} \hline
 & CRA-1 & CRA-2 \\ \hline \hline 
Duration of Stage 1 & $T_{\rm P} = N$ & $T_{\rm P} = N$ \cr \hline
Duration of Stage 2 & $N T_{\rm D} = NM$ 
(fixed) & $D T_{\rm D} = D M$ (varying) \cr \hline
Duration of Feedback & $\frac{\tau}{2}$ & $\tau$ \cr \hline
\end{tabular}
\end{center}	\label{TBL:1}
\end{table}

Note that CRA-1 requires MUD \cite{VerduBook} \cite{ChoiJBook2}
to detect multiple spread signals simultaneously.
On the other hand in CRA-2, since each slot has only one data packet,
a conventional single-user detector can be used.
From this, in terms of data packet detection/decoding,
CRA-2 could be easier to implement than CRA-1.

\section{Throughput Analysis}	\label{S:Thp}

In this section, we consider the throughput 
of the two-stage random access schemes, i.e., CRA-1 and CRA-2.
We first identify possible error 
events and find the throughput.

\subsection{Error Events}

We assume that there is no feedback error from the BS to users
(i.e., Feedbacks 1 and 2 are always successful).

In Stage 1 (for both
CRA-1 and CRA-2), there are the following error events at the BS:
\begin{itemize}
\item False alarm (FA): The BS erroneously detects
a preamble that is not transmitted by any active user. 
\item Missed detection (MD): The BS cannot detect
a preamble transmitted by an active user. 
\item PC: A detected
preamble is 
transmitted by multiple active
users.
\end{itemize}

Note that if each user 
has a unique preamble (as a signature sequence),
there is no preamble collision.
However, if the number of users is large,
it is difficult to manage a set of signature sequences.
Furthermore, if there are new or outgoing users with
unique sequences, the BS needs to update the set
of signature sequences. Therefore, it is convenient to 
have a pool of pre-determined preambles that are shared by users
at the cost of preamble collision.

\subsection{Analysis of CRA-2 with Packet Loss}

Since the analysis of CRA-1 is straightforward once
we analyze CRA-2, we first consider CRA-2 in this subsection.
Throughout this section, we assume that
the active users associated with MD and PC events
drop their packets.

In Stage 2 of CRA-2, for tractable analysis,
we assume that all the transmitted data packets 
are successfully decodable if there is no collision.
In Stage 1, when there are multiple active users
that transmit the same preamble, the BS detects 
the preamble and broadcasts its index through Feedback 1.
Then, all the active users that choose the preamble
transmit their data packets 
in Stage 2, which results in packet collision.
Thus, the packet collision (in Stage 2) is directly related
to preamble collision (in Stage 1).

For convenience, define 
\begin{align}
X_{k,l} =
\left\{
\begin{array}{ll}
1, & \mbox{if active user $k$ chooses preamble $l$} \cr
0, & \mbox{o.w.} \cr
\end{array}
\right.
\end{align}
In addition, let
\begin{align}
\cB = \left\{l: \  \sum_{k=1}^K X_{k,l} \ge 1 \right\} \ \mbox{and} \
\cB_1 = \left\{l: \  \sum_{k=1}^K X_{k,l} = 1 \right\}.
\end{align}
Clearly, $\cB_1 \subseteq \cB$.
Let $B = |\cB|$ and $B_1 = \cB_1$. Then,
$B$ becomes the number of the selected preambles by $K$ active users,
while $B_1$ becomes the number of the preambles,
each of which is selected by only one active user.
Thus, $B \ge B_1$. Let 
$\cB_2 = \cB \setminus \cB_1$ and $B_2 = |\cB_2|$.
Clearly, $B_2 = B - B_1$, which
becomes the number of the preambles, 
each of which is selected
by multiple active users.

\begin{figure}[thb]
\begin{center}
\includegraphics[width=6cm]{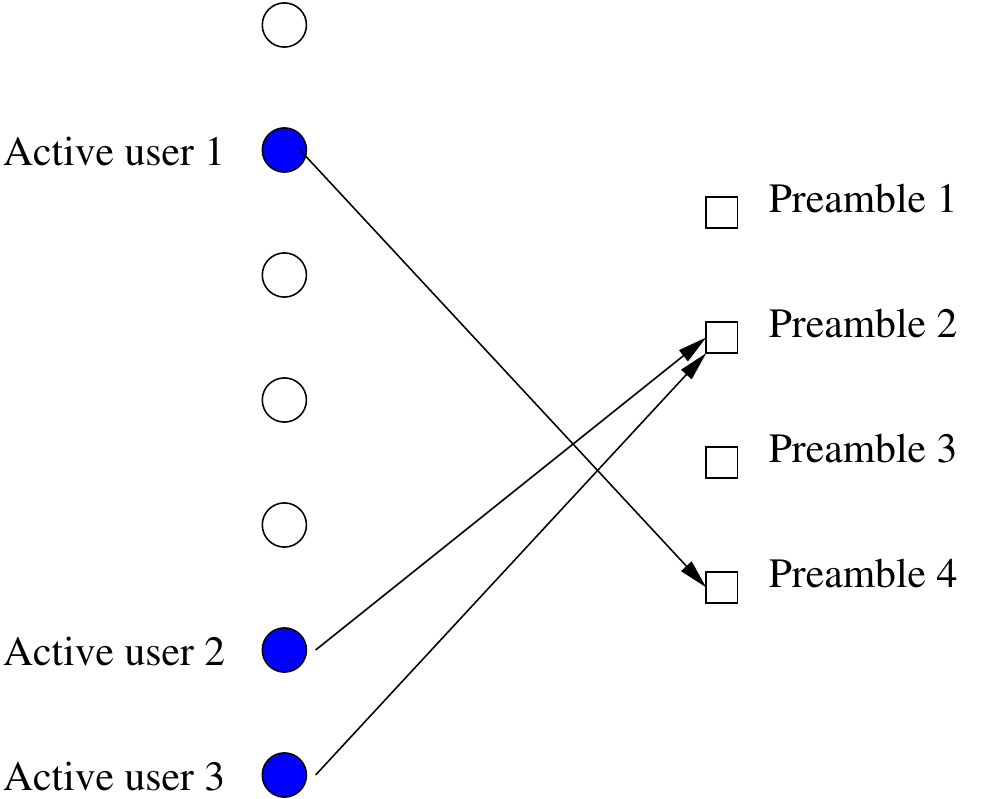}
\end{center}
\caption{There are $K = 3$ active users and $L = 4$ preambles,
where active user 1 
chooses preamble 4 and active users 2 and 3 choose preamble 2.}
        \label{Fig:Xmap}
\end{figure}

In Fig.~\ref{Fig:Xmap},
an example is illustrated 
with $K = 3$ active users and $L = 4$ preambles.
Here, $X_{1,4} =  X_{2,2} = X_{3, 2} = 1$, while
all the other $X_{k,l}$'s are zero.
Clearly, $\cB = \{2, 4\}$ and $\cB_1 = \{4\}$.

At the BS, if FA and 
MD events do not happen, i.e.,
the BS is able to detect all
transmitted preambles,
$D$ becomes $B$.
However, due to FA and MD events, 
$D$ could be different from $B$.
Consider FA events. The number of 
erroneously detected preambles by FA events 
is denoted by $D_3$.
Clearly, $D_3$ cannot be greater than
$L - B$.
Denote by $D_1$ and $D_2$
the numbers of correctly detected 
preambles among those in $\cB_1$ and $\cB_2$, respectively.
Then, we have
$$
D = D_1 + D_2 + D_3.
$$
Among $D$ slots, 
$D_3$ slots will not 
have any data packets 
as there are no active users associated with
them (due to FA events). In addition, each of $D_2$ slots may
have collided packets as there are multiple active users
associated with it (due to PC events).
Consequently, there are only $D_1$ slots, where
each slot has only one data packet from an active user
(i.e., without collision).


Let
\begin{align}
U_l & = \indicator\left( \sum_{k=1}^K X_{k,l} = 1\right) 
\in \{0,1\}\cr
W_l & = \indicator\left( \sum_{k=1}^K X_{k,l} \ge 2 \right)
\in \{0,1\},
	\label{EQ:UW}
\end{align}
where $\indicator(\cdot)$ represents the indicator function.
If we assume that each active user randomly chooses
one of $L$ preambles,
it can be shown that
\begin{align}
\Pr(U_l = 1) 
= \binom{K}{1} \frac{1}{L} 
\left(1 - \frac{1}{L} \right)^{K-1} = 
\frac{K}{L} \left(1 - \frac{1}{L} \right)^{K-1}.
\end{align}
In addition, we have
\begin{align}
\Pr(W_l = 1) 
= 1 - \frac{K}{L} 
\left(1 - \frac{1}{L} \right)^{K-1} - 
\left(1 - \frac{1}{L} \right)^{K}. 
\end{align}
For convenience,
let $\alpha_1 = 
\frac{K}{L} \left(1 - \frac{1}{L} \right)^{K-1}$
and $\alpha_2 = 
\left(1 - \frac{1}{L} \right)^{K}$. 
Then, $\Pr(U_l =1 ) = \alpha_1$
and $\Pr(W_l =1 ) = 1 - \alpha_1 - \alpha_2$.

For tractable analysis, we consider the following assumption.
\begin{itemize}
\item[{\bf A1)}] The $U_l$'s are independent of each other. 
In addition, the $W_l$'s are independent of each other. 
\end{itemize}
The above independent assumption is not valid, but would
be a good approximation for a sufficiently large $L$
with $K \ll L$.
In addition, for convenience, let
\begin{align}
\bar D_i (K) = \uE[D_i \,|\, K],  \ i = 1,2,3.
\end{align}

\begin{mylemma}	\label{L:1}
Suppose that the events of FA and MD
are independent and the probabilities of MD (per 
transmitted preamble) and FA (per untransmitted preamble)
are denoted by $P_{\rm MD}$ and $P_{\rm FA}$,
respectively.
Under the assumption of {\bf A1},
we have
\begin{align}
\bar D_1 (K) & = (1 - P_{\rm MD}) L \alpha_1 \cr
\bar D_2 (K) & = (1 - P_{\rm MD}) L (1 - \alpha_1 - \alpha_2)  \cr
\bar D_3 (K) & = P_{\rm FA} L \alpha_2.
	\label{EQ:L1}
\end{align}
\end{mylemma}
\begin{IEEEproof}
See Appendix~\ref{A:1}.
\end{IEEEproof}

We now consider a steady state analysis using \eqref{EQ:L1}
under the following assumption.
\begin{itemize}
\item[{\bf A2)}] The number of active users, $K$,
follows a Poisson  distribution
with a traffic intensity $\lambda$ (in the number of users per unit time).
\end{itemize}
Let $D(t)$ denote the number of detected preambles in session $t$
and denote by $Y(t)$ the length of session $t$.
Then, we have
\be
Y(t) = T_{\rm P} + \tau + T_{\rm D} D(t),
	\label{EQ:YT}
\ee
where $\tau$ represents the total length of Feedbacks 1 or 2.
For convenience, let $\tilde T_{\rm P} = T_{\rm P} + \tau$.
Then, the number of the active users in slot $t+1$ follows
the following distribution:
\be
\Pr(K (t+1) = k) = 
\frac{ (\lambda Y(t) )^k}{k!} e^{-\lambda Y(t)}.
\ee
We assume that the mean of $Y(t)$ exists and is denoted by 
$\bar Y = \uE[Y(t)]$ and let $\beta_2 = \lambda \bar Y$.
Furthermore, in the steady state ($t \to \infty$), we assume that
$K$ has the following steady-state distribution:
\be
\Pr(K = k) = 
\frac{ \beta_2^k}{k!} e^{-\beta_2}.
	\label{EQ:Kbeta_2}
\ee
In the following result, we obtain an expression for $\beta_2$ if
$\bar Y$ exists.

\begin{mylemma}	\label{L:2}
With the steady-state distribution of $K$ in \eqref{EQ:Kbeta_2},
$\beta_2$ is given by
\begin{align}
\beta_2 & = L \left(c_1 +  \uW (- c_2 e^{-c_1} ) \right) 
	\label{EQ:L2} \\
\bar D & = L \left(1 - P_{\rm MD} +
\frac{\uW (-c_2 e^{-c_1}) }{\lambda T_{\rm D}} 
	\label{EQ:L2b}
\right).
\end{align}
where $\uW (\cdot)$ denotes the Lambert W function\footnote{The
Lambert function is the inverse function of $f(x) = xe^x$,
i.e., $x = \uW (y)$, where $y = x e^x$.} 
and
\begin{align}
c_1 & = \lambda \left(\frac{\tilde T_{\rm P}}{L}+ 
T_{\rm D} (1-P_{\rm MD}) \right) \cr
c_2 & = \lambda T_{\rm D} (1 - P_{\rm MD} - P_{\rm FA}).
\end{align}
\end{mylemma}
\begin{IEEEproof}
See Appendix~\ref{A:2}.
\end{IEEEproof}

We can also obtain 
the mean of $D_1$ as follows:
\begin{align}
\bar D_1 & = \uE[D_1 (K)] = \sum_{k=1}^\infty \bar D_1 (k) 
\frac{\beta_2^k}{k!} e^{-\beta_2} \cr
& = (1- P_{\rm MD}) \beta_2 e^{- \frac{\beta_2}{L}}.
	\label{EQ:bD1}
\end{align}
Thus, the average number of
successfully transmitted packets per session
(or the ratio of 
the average number of
successfully transmitted packets to the average
session length) is given by
\begin{align}
\eta_2 = \frac{\bar D_1}{\bar Y} =
\frac{(1- P_{\rm MD}) \beta_2 e^{- \frac{\beta_2}{L}}}{\tilde T_1
+ \bar D T_2}  = \lambda
(1- P_{\rm MD}) e^{- \frac{\beta_2}{L}},
	\label{EQ:eta2}
\end{align}
which is the throughput of CRA-2.

Note that we have assumed that the error probabilities,
$P_{\rm MD}$ and $P_{\rm FA}$, are constants
(i.e., independent of $K$). This assumption will be discussed
in Subsection~\ref{SS:EP}.

\subsection{Analysis of CRA-1 with Packet Loss}

In this subsection, we focus on the throughput
of CRA-1.

Let $Z = T_{\rm P} + \frac{\tau}{2} + N T_{\rm D}$.
Then, the probability that $K = k$ in CRA-1 is given by
$\Pr(K = k) = \frac{\eta_1^k}{k!} e^{-\beta_1}$,
where $\beta_1 = \lambda Z$,
which is seen as the average number of active users.
In CRA-1, as mentioned earlier,
we assume that if $K \ge N$
(i.e., the number of active users is greater
than the spreading gain), MUD fails.
Thus, the average number of successfully recovered
packets becomes
\begin{align}
\bar K_1 
& = \sum_{k=1}^{N -1}\bar D_1 (k) \Pr(K = k) \cr
& =
(1- P_{\rm MD}) \beta_1 e^{-\beta_1} \sum_{k=0}^{N-2}
\frac{\left(\beta_1 \left(1-\frac{1}{L} \right)\right)^k}{k!} \cr
& =  
(1- P_{\rm MD}) \beta_1 e^{- \beta_1}
\frac{ \Gamma \left(N-1, \beta_1\left(1-\frac{1}{L} \right)
\right)}{(N-1)!},
	\label{EQ:bK1}
\end{align}
where $\Gamma(n,x) = \int_x^\infty t^{n-1} e^{-t} dt$ represents
the upper incomplete gamma function.
Then, the throughput of CRA-1 
(as the ratio of the average number of successfully transmitted
packets to the session length)
is given by
\begin{align}
\eta_1 
= \frac{\bar K_1}{Z} 
= 
\lambda (1- P_{\rm MD})  e^{- \beta_1}
\frac{ \Gamma \left(N-1, \beta_1\left(1-\frac{1}{L} \right)
\right)}{(N-1)!}.
	\label{EQ:eta_1}
\end{align}
For convenience, let $V$ be a Poisson random variable
with mean $\beta_1 \left(1 - \frac{1}{L} \right)$.
Then, it can be shown that
\be
\eta_1 
= \lambda (1 - P_{\rm MD}) e^{-\frac{\beta_1}{L}}
\Pr(V \le N-2).
	\label{EQ:eta11}
\ee
From \eqref{EQ:eta2}
and \eqref{EQ:eta11}, we can see that
if $\beta_1 \approx \beta_2$ (which is the case
that the average duration of Stage 2 in CRA-2 
is equal to that in CRA-1 or $\bar D \approx N$),
$\eta_2$ might be larger than $\eta_1$ as 
$\Pr(V \le N-2) \le 1$ and the difference
may increase if 
$\uE[V] -(N-2) = \beta_1 \left(1 - \frac{1}{L} \right) - (N-2)$
is positive and increases.

Note that when multichannel ALOHA is used,
in \eqref{EQ:bK1}, $L$ is to be replaced with $N$.
In addition, since all the channels are orthogonal,
the BS can decode up to $N$ packets when each packet is transmitted
with a different preamble.
Thus, the throughput of multichannel ALOHA becomes
\begin{align}
\eta_{\rm ma} 
 = 
\lambda (1- P_{\rm MD})  e^{- \beta_1}
\frac{ \Gamma \left(N, \beta_1\left(1-\frac{1}{N} \right)
\right)}{N!}.
	\label{EQ:eta_ma}
\end{align}
For a sufficiently large $N$, we can see that
$\eta_1 > \eta_{\rm ma}$ for $L > N$.

\section{Other Issues}	\label{S:OI}

In this section, we discuss two key issues 
that are not considered for
the throughput analysis in Section~\ref{S:Thp}.

\subsection{Stability Issue with Re-Transmissions}

In Section~\ref{S:Thp},
we do not consider re-transmissions of  unsuccessful 
data packets.
Since there are multiple preambles, an active user
with unsuccessful data packet can immediately transmit
another randomly chosen preamble in the next session,
which can be seen as fast retrial \cite{YJChoi06}.
For CRA-1, fast retrial was considered in 
\cite{Choi_CRA18} with its stability analysis when the access
probability is controlled.
In this subsection, we briefly discuss stability issues with fast retrial
for CRA-2.

Let $t$ be the index for sessions. 
Thus, $D_i (t)$ represents $D_i$ in session $t$.
After session $t$,
the number of the backlogged users due to preamble collision
and MD events is given by
\be
Z(t) = K(t)  - D_{1} (t),
\ee
where $K(t)$ denotes the number of users at session $t$.
Thus, 
if the active users with unsuccessfully
transmitted packets attempt to transmit preambles in the next
session based on fast retrial,
we have
\begin{align}
\uE[K(t+1)\,|\, D(t)] 
& = \lambda Y(t)  + Z(t)  \cr
& = \lambda \left(T_{\rm P} + \tau + T_{\rm D} D(t) \right)  + Z(t).
\end{align}
Let
$\uE[D(t) \,|\, K(t)] = \sum_{i=1}^3 \bar D_i (K(t))$.
Since
\begin{align}
\uE[K(t+1)\,|\, K(t)] 
& = \lambda \left(\tilde T_{\rm P} + T_{\rm D} \uE[D(t)\,|\, K(t)]  \right) \cr
& \ + K(t)  - \uE[D_1 (t)\,|\, K(t)] \cr
& = \lambda \left(\tilde T_{\rm P} + T_{\rm D} \left(
\sum_{i=2}^3 \bar D_i (K(t)) \right) \right) \cr
& \ + K(t)  - (1 - \lambda T_{\rm D}) \uE[D_1 (t)\,|\, K(t)], \ \quad 
\end{align}
the drift \cite{Kelly_Yudovina} becomes
\begin{align}
\delta (K) & = \uE[K(t+1)\,|\, K(t) = K]  - K \cr
& = \lambda \left(\tilde T_{\rm P} + T_{\rm D} \left(
\sum_{i=2}^3 \bar D_i (K) \right) \right)  \cr
& \quad - (1 - \lambda T_{\rm D}) \bar D_1(K). 
\end{align}
From \eqref{EQ:L1},
it can be shown that $\lim_{K \to \infty} \bar D_1 (K) = 
\lim_{K \to \infty} \bar D_3 (K) = 0$ and 
$\lim_{K \to \infty} \bar D_2 (K) = (1-P_{\rm MD}) L$.
Then, it 
can be shown that $\delta (K) > 0$ for $K \ge K_0$, where $K_0$ is finite.
This implies that CRA-2 becomes unstable if $K$ is sufficiently large,
which results from a number of unsuccessful packets that are to be
re-transmitted.
To avoid this problem, 
unsuccessful packets can be dropped as in Section~\ref{S:Thp} or
access control 
schemes can be considered, which might be studied in the future.

\subsection{Error Probabilities}	\label{SS:EP}

In CRA, for MUD, greedy CS algorithms \cite{Eldar12}
are used
to detect all active users (actually their preambles)
under the assumption that the number of active users
is sufficiently small, e.g., \cite{Applebaum12} \cite{Schepker13}.
In \cite{Seo19}, LASSO \cite{Tibshirani96}
is applied to MUD in order to detect transmitted preambles.
In general, the detection performance of 
transmitted preambles depends on the algorithm used for MUD.
In other words, $P_{\rm MD}$ and $P_{\rm FA}$ depend on the algorithm used.
Thus, in this subsection, 
in order to avoid the performance dependency
on a particular algorithm,
we mainly focus on optimal MUD and consider 
$P_{\rm MD}$ and $P_{\rm FA}$ based on
the maximum likelihood (ML)\footnote{In general,
for ML detection, an exhaustive search is used. In this case,
the complexity grows exponentially with $L$, which makes
the use of ML detection impractical. However, we consider
ML detection to see achievable performance.}
criterion.

Provided that there are $K$ active users,
it might be unlikely to detect more than $K+1$ users
in the event of FA or less than $K-1$ users
in the event of MD.
Thus, for tractable analysis, we only consider the case that 
the BS detects $K+1$ transmitted preambles 
for FA events. 
Likewise,
for MD events, the case that 
the BS detects $K-1$ transmitted preambles is 
studied. 
In addition, for simplicity,
we assume that there is no preamble collision.
Thus, $B = K$ in this subsection.

It is assumed that all the preambles are normalized
as $||\bpsi_l||_2 = 1$.
Let the support set of $\bs$ be 
\be
\cS = {\rm supp}(\bs) = \{l(1), \ldots, l(K)\}.
\ee
In addition, for MD events,
define a vector $\bs^\prime_k$ as follows:
\be
[\bs - \bs^\prime_k]_l = 
\left\{
\begin{array}{ll}
\sqrt{P_k} h_k, & \mbox{if $l = l(k)$} \cr
0, & \mbox{o.w.} \cr
\end{array}
\right.
\ee
Clearly, $\bs^\prime_k$ is a $(K-1)$-sparse
vector that does not include the signal from the $k$th active user.
From \eqref{EQ:by},
if the
ML detector is
used with known CSI (i.e., all users' channel coefficients,
$h_k$'s),
it can choose $\bs^\prime_k$ 
with the following probability:
\begin{align}
P_k^\prime 
& = \Pr(||\by - \bPsi \bs||^2 > ||\by - \bPsi \bs_k^\prime||^2 ) \cr
& = \Pr(||\bn||^2 > ||\bpsi_{l(k)} \sqrt{P_k} h_k  + \bn||^2 ) \cr
& = \cQ \left( \sqrt\frac{ ||\bpsi_{l(k)}||^2 P_k|h_k|^2 }{2 N_0} \right)
= \cQ \left( \sqrt\frac{P_k|h_k|^2 }{2 N_0} \right). \ \ 
	\label{EQ:Pk}
\end{align}
The probability in \eqref{EQ:Pk} is to be seen as a lower-bound,
because $\{h_k\}$ (i.e., the CSI)
is to be known, which has to be estimated in a CS-based MUD.

As mentioned earlier, since
we only consider the events of MD with only one user
not detected, using the union bound, the probability of MD event
(per used preamble)
becomes
\be
P_{\rm MD} \le \frac{1}{K} \sum_{k=1}^K P_k^\prime.
	\label{EQ:PMD}
\ee

For FA events, consider a virtual (active) user, say user $K+i$, $i 
\in \{1, \ldots, L-K\}$,
that sends preamble $\bpsi_{l(K+i)}$,
where $l(K+i) \in \cS^c$. For each virtual user, 
consider a $(K+1)$-sparse vector,
denoted by
$\bs^{\prime \prime}_{K+i}$, that satisfies
\be
[\bs^{\prime \prime}_{K+i} - \bs]_l = 
\left\{
\begin{array}{ll}
\sqrt{P_{K+i}} h_{K+i}, & \mbox{if $l = l(K+i)$} \cr
0, & \mbox{o.w.} \cr
\end{array}
\right.
	\label{EQ:vu}
\ee
Here, $h_{K+i}$ is the channel coefficient
for a virtual user (say user $K+i$) and $P_{K+i}$
represents its transmit power.
Note that although user $K+i$ is not an 
active user, for the FA event, we consider a hypothesis that this user 
is incorrectly detected as an active user. To this end,
this user's channel coefficient and transmit power are assumed as in 
\eqref{EQ:vu}.
Then, the probability that the BS incorrectly chooses
$\bs_{K+i}^{\prime \prime}$ 
is 
\begin{align}
P_{k+i}^{\prime \prime}
& = \Pr(||\by - \bPsi \bs ||^2 > 
||\by - \bPsi \bs_{K+i}^{\prime\prime}||^2 ) \cr
& = \cQ
\left( \sqrt\frac{ P_{K+i}|h_{K+i}|^2 }{2 N_0} \right).
\end{align}
Using the union bound, the probability of FA event (per unused
preamble)
becomes
\begin{align}
P_{\rm FA} \le
\frac{1}{L - K} \sum_{i=1}^{L-K} P_{k+i}^{\prime \prime}.
	\label{EQ:PFA}
\end{align}
Although we do not show
the probabilities of MD and FA events
with more than 1 user difference, they are much lower
than those with 1 user difference in 
\eqref{EQ:PMD} and \eqref{EQ:PFA}. Note that
more detailed analysis is required
if the correlation of preambles is high or SNR is low, 
which could be a further research issue to be studied in the future.

As we can see in \eqref{EQ:PMD} and \eqref{EQ:PFA},
$P_{\rm MD}$ and $P_{\rm FA}$
are mainly dependent on the SNR,
$\frac{P_k |h_k|^2}{N_0}$. 
With a power control policy to compensate 
fading, we may have $P_k |h_k|^2 = P_{\rm rx}$,
where $P_{\rm rx}$ represents the effective receive power. 
With a sufficiently high SNR,
$P_{\rm MD}$ and $P_{\rm FA}$ can be sufficiently low
and independent of $K$.
This justifies the use of constant error probabilities,
$P_{\rm MD}$ and  $P_{\rm FA}$, in finding
the throughput in Section~\ref{S:Thp}.

Note that 
although $P_{\rm MD}$ and $P_{\rm FA}$ are sufficiently
low, the probability that 
the transmitted preambles are not correctly detected by
MUD (i.e., the error probability) can be high.
To see this,
let $P_{\rm MD} = P_{\rm FA} = \epsilon \ll 1$.
Then, assuming that MD and FA events are independent,
the error probability becomes
\begin{align}
P_{\rm err} & = \Pr(\hat \cS \ne \cS) \cr
& = 1 - (1- P_{\rm FA})^{L-K} (1- P_{\rm FA})^{K} \cr
& \approx 1 - e^{-(L-K) \epsilon} e^{-K \epsilon} 
= 1 - e^{-L\epsilon},
\end{align}
where $\hat \cS$ represents the estimated support set by MUD.
For example, if $\epsilon = 0.01$ and $L = 310$, 
$P_{\rm err}$ becomes $0.955$,
which indicates that
even if $P_{\rm err}$ can be high,
the probability of individual MD or FA event (per used preamble)
can be low.

\section{Simulation Results}	\label{S:Sim}

In this section, we present simulation results for the throughput
and the average length of Stage 2 of CRA-2, 
and compare them with the theoretical ones
in \eqref{EQ:eta2} and \eqref{EQ:bD1}, respectively. 
For convenience, let
$T = T_{\rm P} + T_{\rm D} = N+M$ (in unit symbol
duration\footnote{The unit symbol duration
is the inverse of the bandwidth if the Nyquist sampling rate is used.
All the lengths of sequences (e.g.,
$N$, $M$, $T$, and so on) and feedback duration (e.g., $\tau$) 
are given in unit symbol duration.}), 
which is the transmission
time required to send one packet with one preamble by one active user.
In addition, let $\lambda_T = \lambda T$ be the normalized
traffic intensity. That is, if $\lambda_T = 1$,
it is expected to have one packet is generated over $T$ on average.
The throughput and the traffic intensity will be normalized\footnote{The
normalized throughput is the average number of successfully
recovered packets per $T = N + M$. That is, the normalized throughput is
$T \eta_1$ and $T \eta_2$ for CRA 1 and CRA 2, respectively.}
in most cases in this section.
Note that from \eqref{EQ:PMD} and \eqref{EQ:PFA},
in this section, we assume that $P_{\rm FA} = P_{\rm MD}$.

\begin{figure}[thb]
\begin{center}
\includegraphics[width=\figwidth]{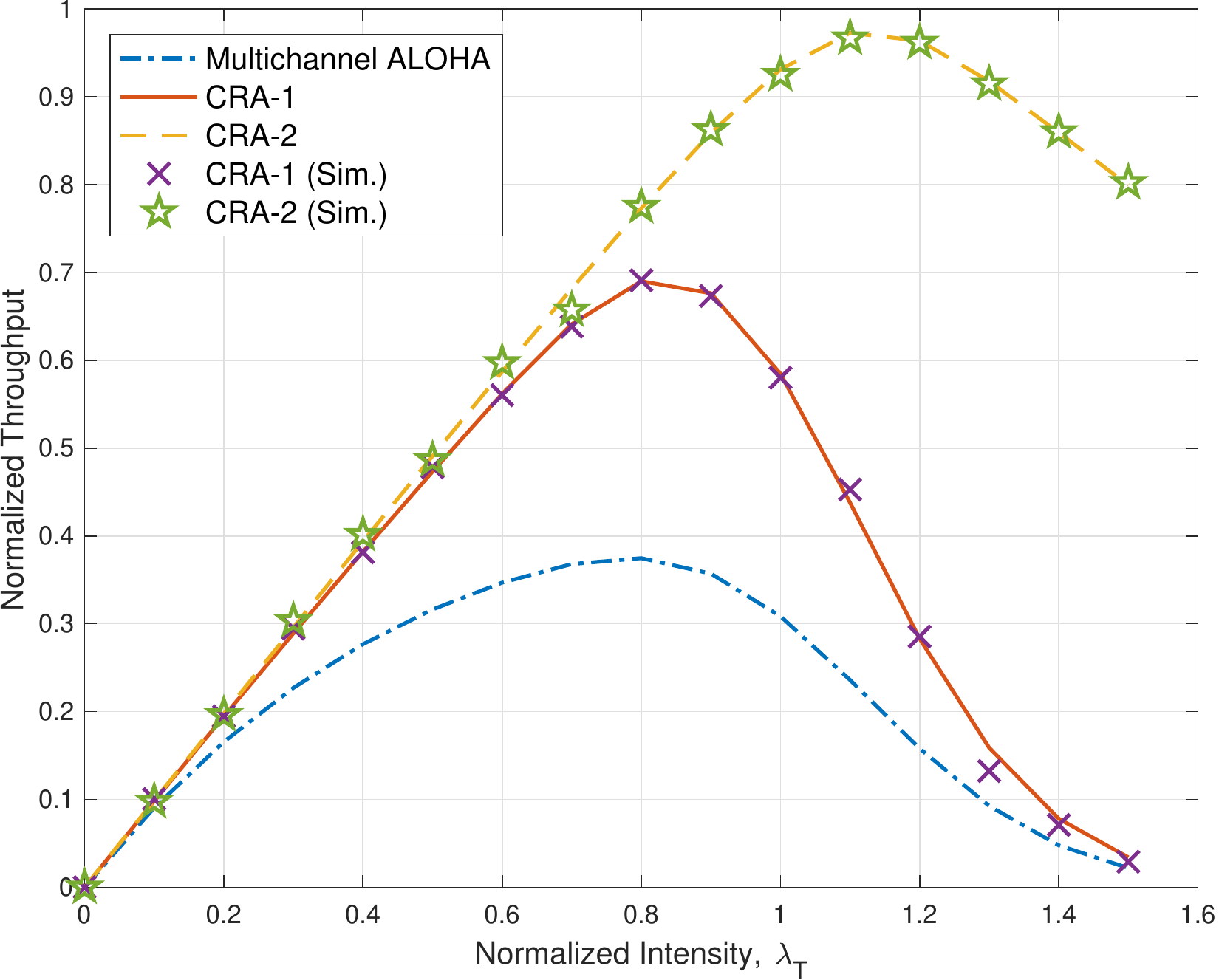} \\
(a) \\
\includegraphics[width=\figwidth]{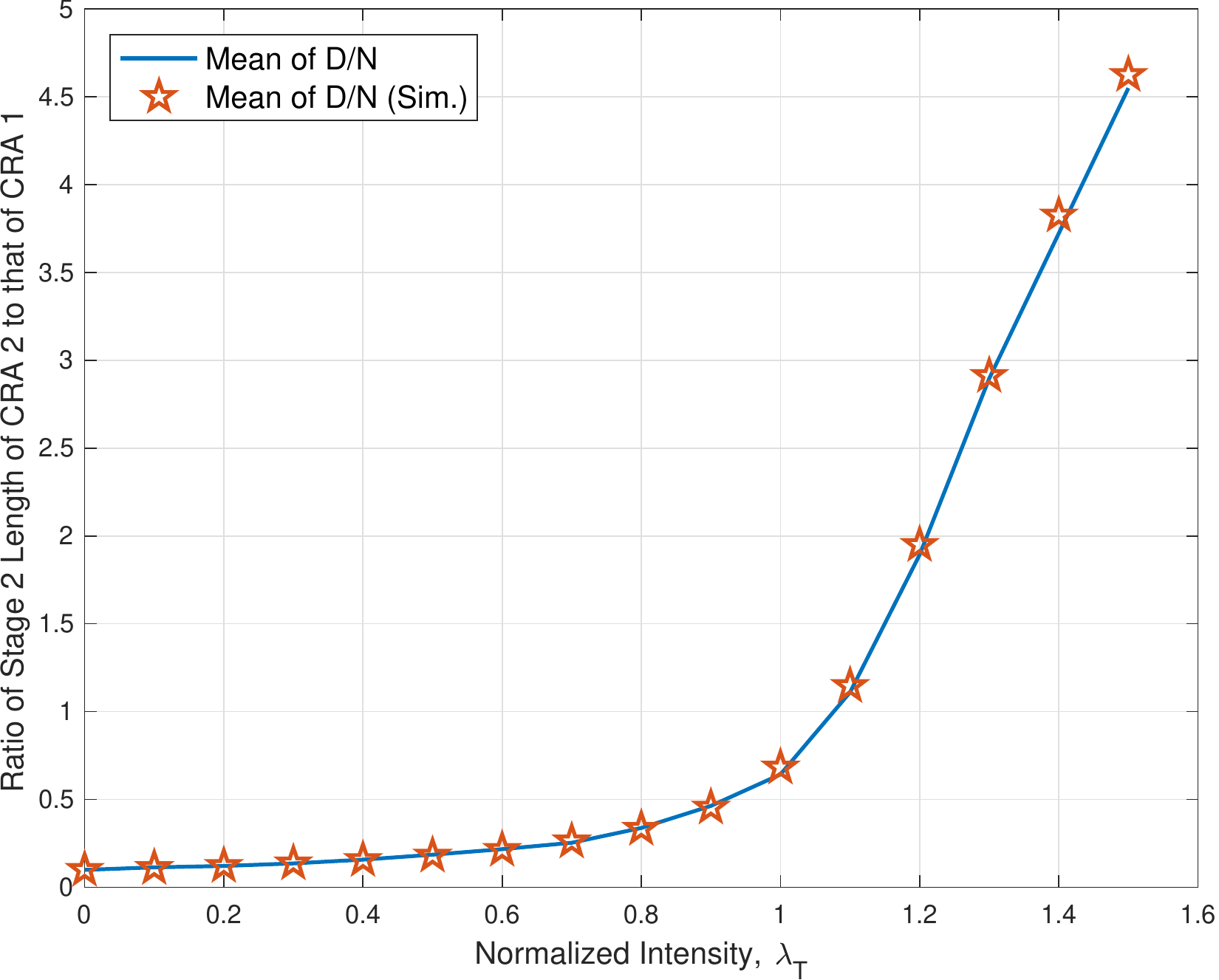} \\
(b) \\
\end{center}
\caption{Performance for various normalized traffic intensity $\lambda_T$
with $N = 31$, $M = 256$, $\tau = 4$, $L  = 10 N$, and 
$P_{\rm FA} = P_{\rm MD} = 0.01$: (a) Normalized throughput,
$T \eta_i$, $i \in \{1, 2, {\rm ma}\}$; 
(b) the ratio of the average length of Stage 2, $\bar D M$, of CRA-2
to that of CRA-1, $NM$.}
        \label{Fig:plt_s1}
\end{figure}

Fig.~\ref{Fig:plt_s1} shows the performance 
for various normalized traffic intensity $\lambda_T$
with $N = 31$, $M = 256$
(if binary signaling is used,
a data packet can transmit 32 bytes), $\tau = 4$, $L  = 10 N$, and 
$P_{\rm FA} = P_{\rm MD} = 0.01$.
The throughput curves of the three different schemes,
CRA-1, CRA-2, and multichannel ALOHA, 
are shown in Fig.~\ref{Fig:plt_s1} (a).
In Fig.~\ref{Fig:plt_s1} (b),
the
ratio of the average length of Stage 2, $\bar D M$, of CRA-2
to that of CRA-1, $NM$
is shown as a function of $\lambda_T$.
Clearly, CRA-1 has
a higher throughput than
multichannel ALOHA by a factor of (nearly) 2,
as known in \cite{Choi_CRA18}.
More importantly, we can see that 
the throughput of CRA-2 is higher than that of CRA-1 and it
can be close 1 thanks to the adaptive length of Stage 2.
If the number of active users
is sufficiently small, CRA-2 may have a shorter
length of Stage 2 than CRA-1 (i.e., $D_1 +D_2 + D_3 < N$),
which results in a higher throughput (than that of CRA-1).
As shown in Fig.~\ref{Fig:plt_s1} (b),
we have $\bar D < N$ if $\lambda_T \le 1$.


\begin{figure}[thb]
\begin{center}
\includegraphics[width=\figwidth]{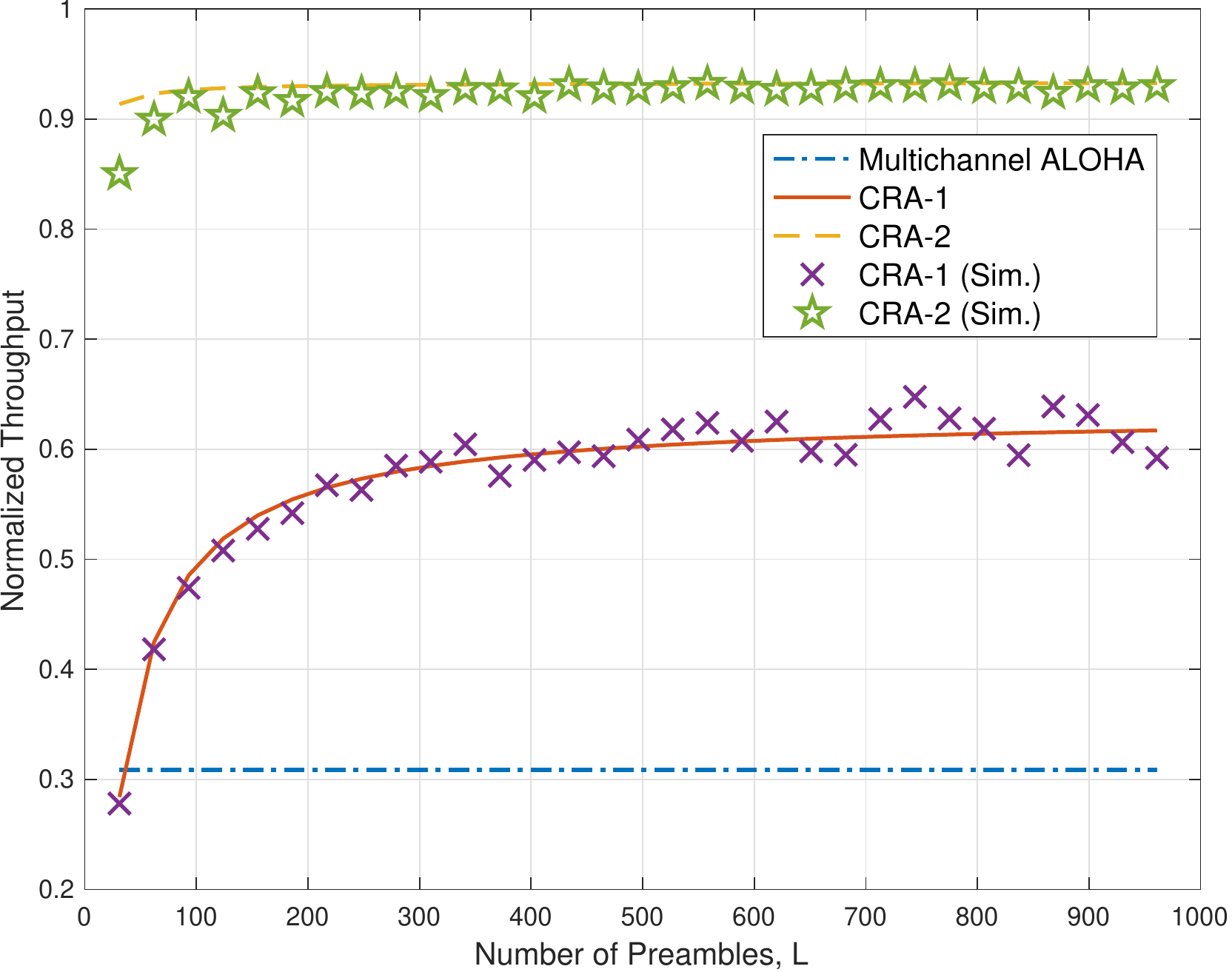} \\
(a) \\
\includegraphics[width=\figwidth]{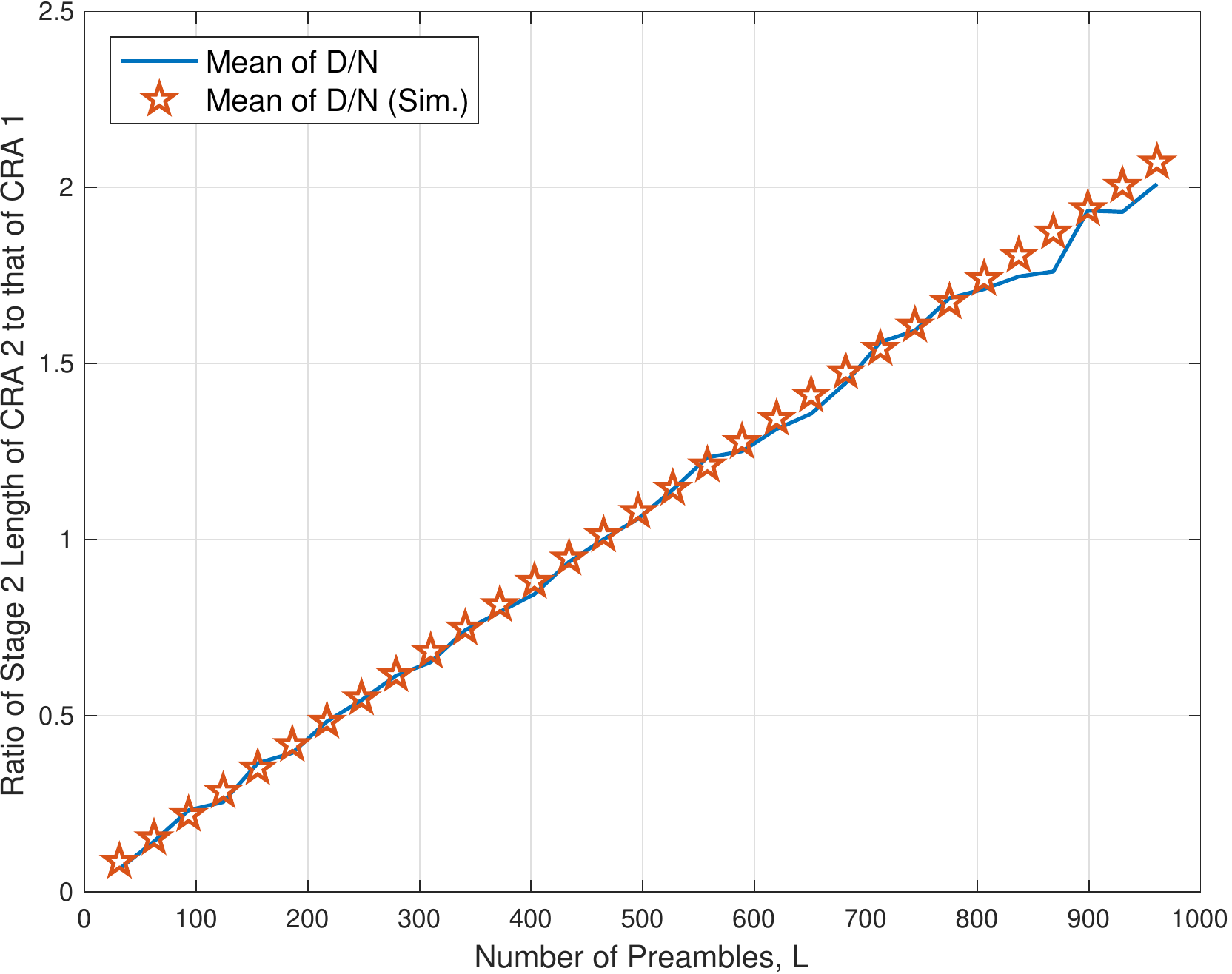} \\
(b) \\
\end{center}
\caption{Performance for various numbers of preambles, $L$,
with $\lambda_T = 1$, $N = 31$, $M = 256$, $\tau = 4$, and 
$P_{\rm FA} = P_{\rm MD} = 0.01$: (a) Normalized throughput,
$T \eta_i$, $i \in \{1, 2, {\rm ma}\}$; 
(b) the ratio of the average length of Stage 2, $\bar D M$, of CRA-2
to that of CRA-1, $NM$.}
        \label{Fig:plt_s2}
\end{figure}

We show performance for various numbers of preambles, $L$,
with $\lambda_T = 1$, $N = 31$, $M = 256$, $\tau = 4$, and 
$P_{\rm FA} = P_{\rm MD} = 0.01$ in Fig.~\ref{Fig:plt_s2}.
It is shown that the theoretical results agree
with simulation results for a wide range of $L$
(except that $L$ is small\footnote{As mentioned earlier,
the assumption of {\bf A1)} is reasonably when 
$L \gg K$. Thus, if $L$ is small, 
$\eta_2$ in
\eqref{EQ:eta2} may differ from simulation results.}
in CRA-2).
In general, as the number of preambles, $L$, increases,
a higher throughput is achieved in both CRA-1 and CRA-2.
However, the throughput becomes saturated once $L$ is sufficiently
large.
This indicates that it is not necessary to have a large 
number of preambles. For example, if $L = 10 N$ might be sufficient
to have a reasonable performance in terms of throughput.
In Fig.~\ref{Fig:plt_s2} (b),
it is also shown that the average length of
CRA-2 increases with $L$. If $L \le 450$, we see that
the average length of CRA-2 is shorter than that of CRA-1. 
Thus, with not too large $L$, 
CRA-2 can provide a higher throughput and a shorter delay than
CRA-1.

\begin{figure}[thb]
\begin{center}
\includegraphics[width=\figwidth]{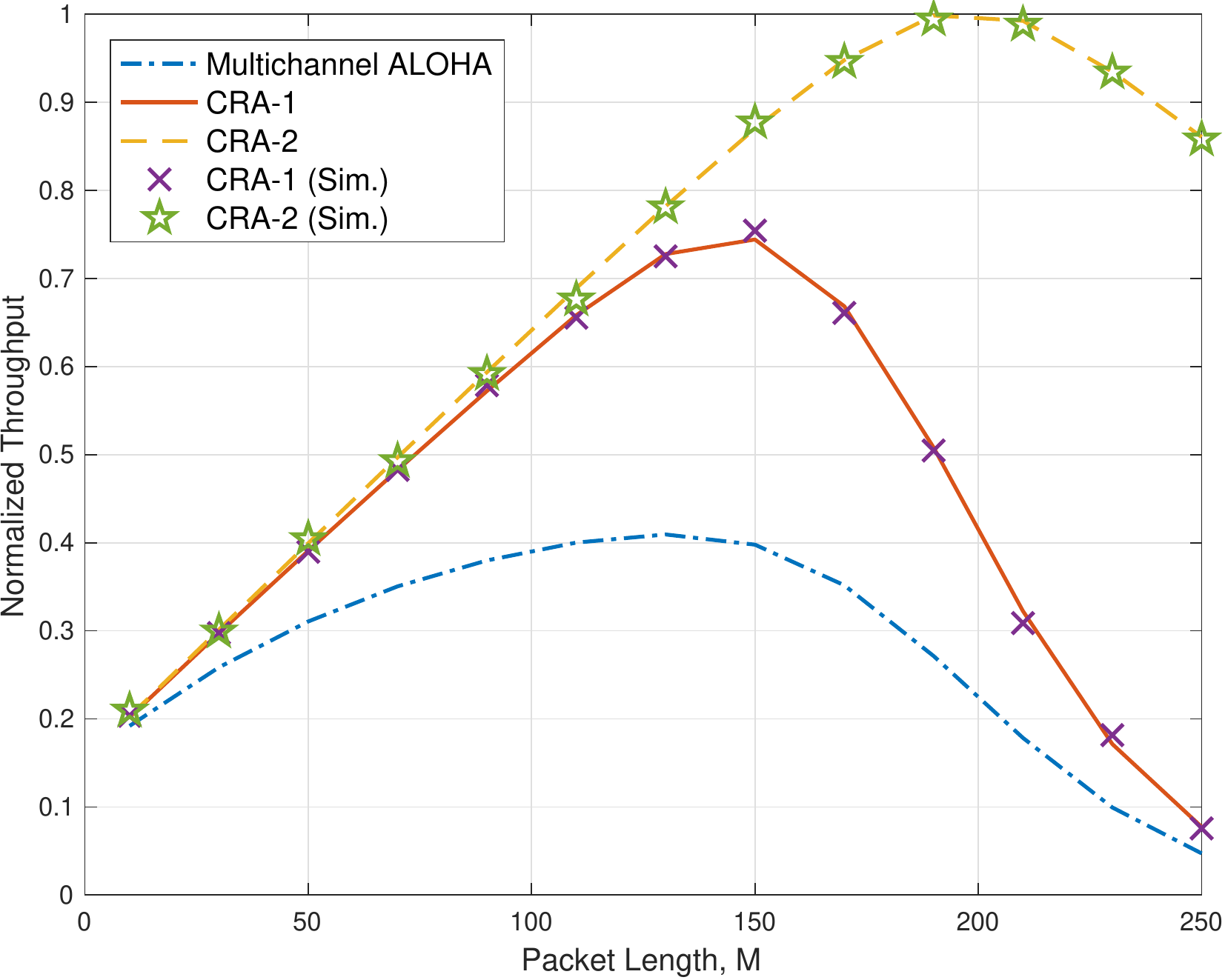} \\
(a) \\
\includegraphics[width=\figwidth]{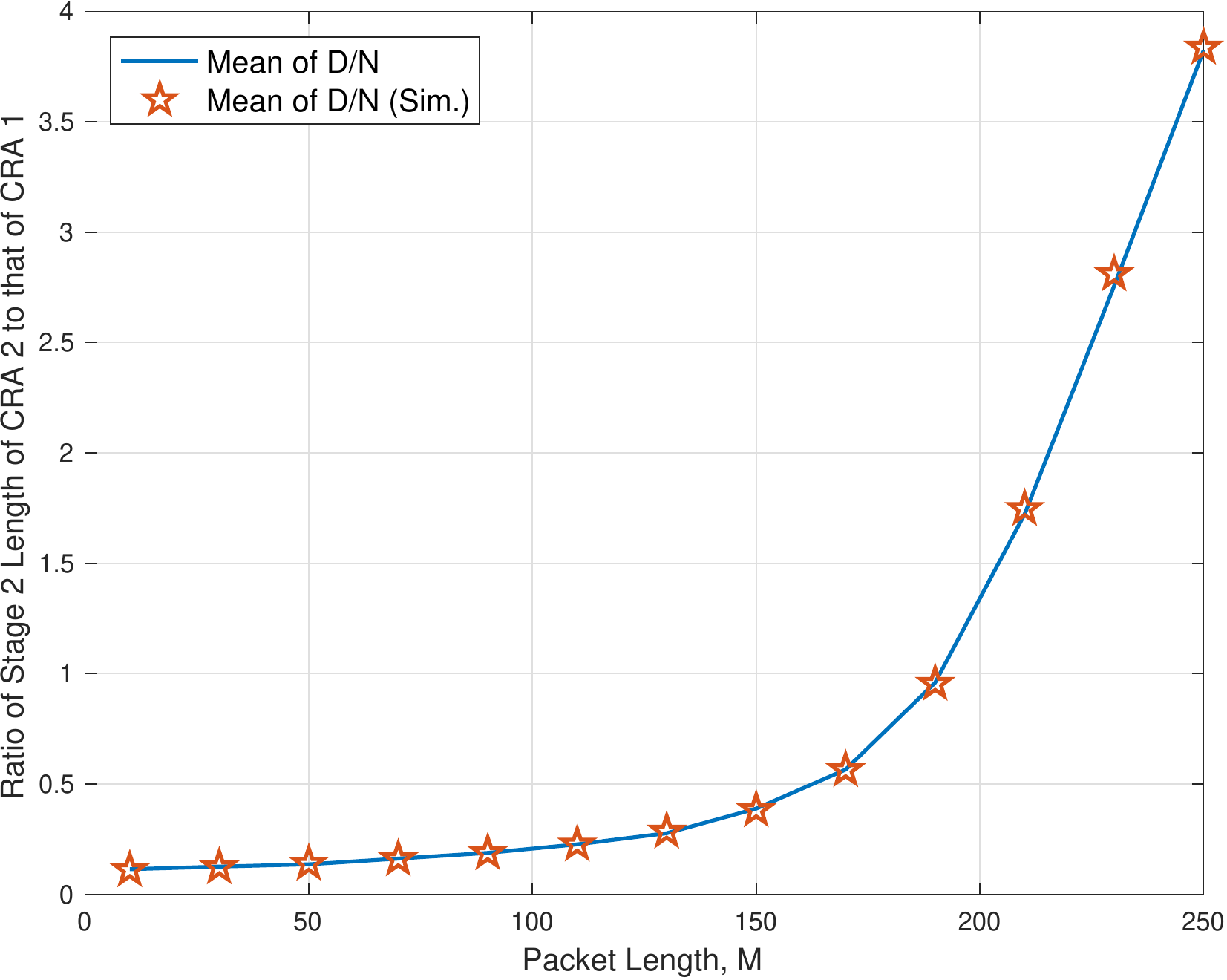} \\
(b) \\
\end{center}
\caption{Performance for various lengths of packets, $M$,
with $\lambda = \frac{1}{200}$, $N = 31$, $L = 10 N$, $\tau = 4$, and 
$P_{\rm FA} = P_{\rm MD} = 0.01$: (a) Normalized throughput,
$T \eta_i$, $i \in \{1, 2, {\rm ma}\}$; 
(b) the ratio of the average length of Stage 2, $\bar D M$, of CRA-2
to that of CRA-1, $NM$.}
        \label{Fig:plt_s3}
\end{figure}

In Fig.~\ref{Fig:plt_s3}, we set $\lambda$ to $\frac{1}{200}$
and show the performance
for various lengths of packet, $M$,
with  $N = 31$, $L = 10 N$, $\tau = 4$, and 
$P_{\rm FA} = P_{\rm MD} = 0.01$.
It is shown that the throughput increases with $M$
and then decreases in all the random access schemes.
Thus, in each scheme, there is a best length of data packet
that maximizes the throughput.
It is shown in
Fig.~\ref{Fig:plt_s3} (b) that
$M$ should not be too long to avoid a long delay
and a low throughput in CRA-2.

\begin{figure}[thb]
\begin{center}
\includegraphics[width=\figwidth]{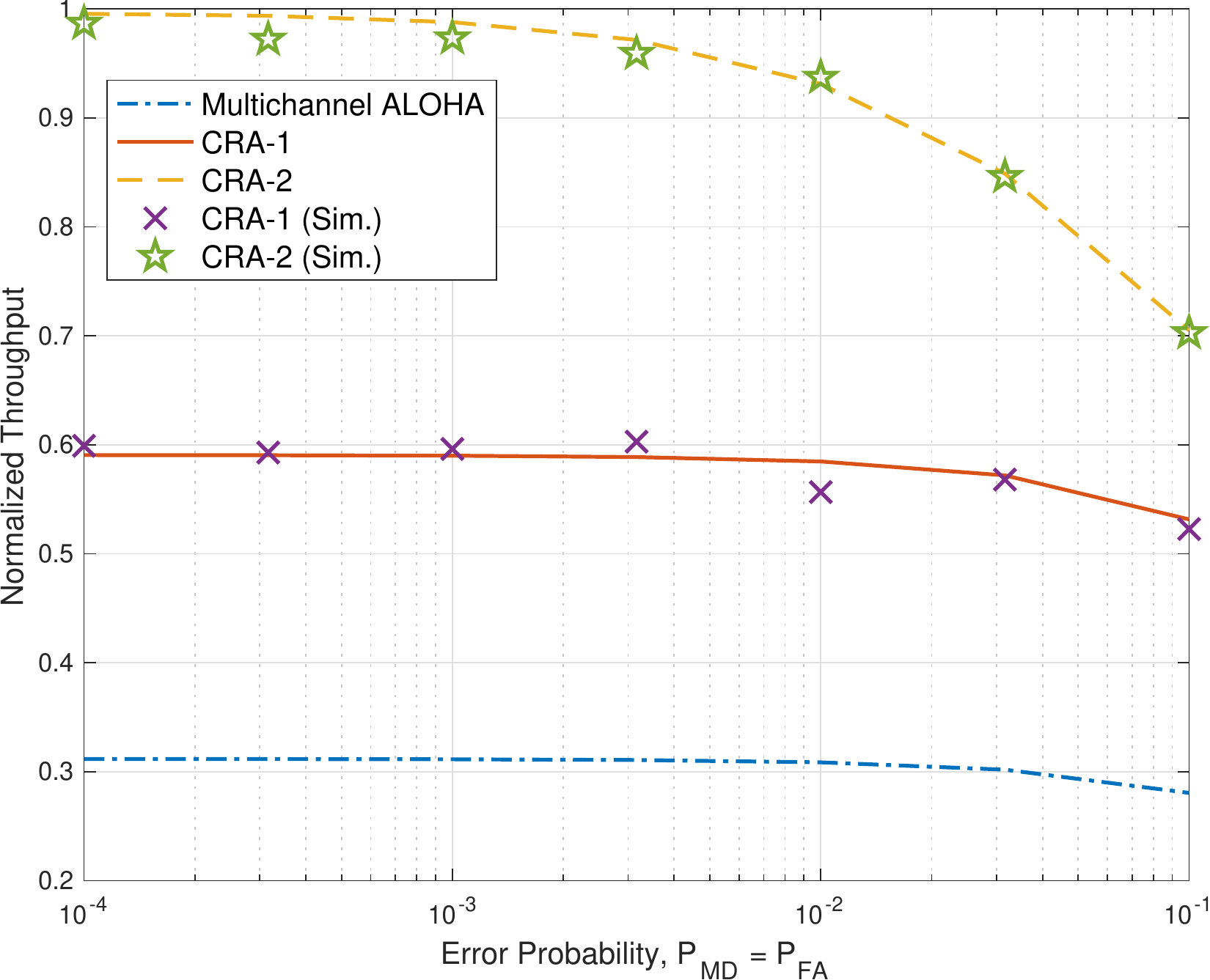} \\
(a) \\
\includegraphics[width=\figwidth]{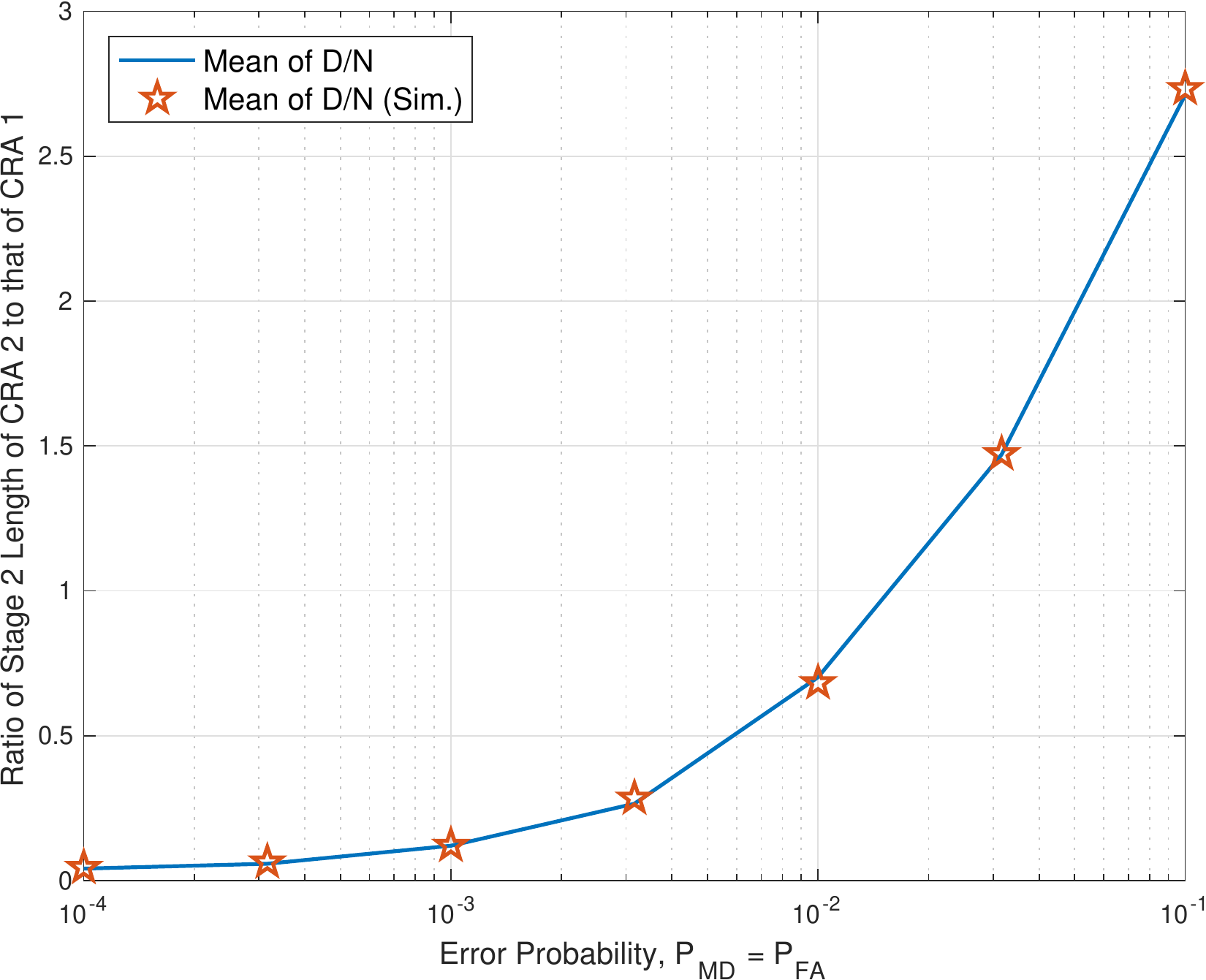} \\
(b) \\
\end{center}
\caption{Performance for various probabilities
of MD and FA
with $\lambda_T = 1$, $N = 31$, $L = 10 N$, $\tau = 4$, and 
$M = 256$: (a) Normalized throughput,
$T \eta_i$, $i \in \{1, 2, {\rm ma}\}$; 
(b) the ratio of the average length of Stage 2, $\bar D M$, of CRA-2
to that of CRA-1, $NM$.}
        \label{Fig:plt_s4}
\end{figure}

The impact of the probabilities of MD and FA
on the performance is shown in
Fig.~\ref{Fig:plt_s4}
with $\lambda_T = 1$, $N = 31$, $L = 10 N$, $\tau = 4$, and 
$M = 256$.
It is shown that CRA-2 has a more 
throughput degradation than other schemes as 
$P_{\rm MD} = P_{\rm FA}$ increases thanks to
slots generated by FA events, (i.e., $D_3$ slots).
However, as long as $P_{\rm FA} = P_{\rm MD}$ is sufficiently
low, CRA-2 outperforms the others. 
In addition, as shown in 
Fig.~\ref{Fig:plt_s4} (b), the average length of Stage 2
increases with $P_{\rm FA} = P_{\rm MD}$.
Thus, to avoid a long delay, 
it is necessary to keep $P_{\rm FA} = P_{\rm MD}$ low.

\section{Concluding Remarks}	\label{S:Conc}

In a number of IoT applications,
each active IoT device or sensor may
have a short data packet
(of a few ten bytes) and 
one short message delivery would be required at each access
to uplink. Thus, we considered CRA-based random access
schemes when active devices or sensors
have one data packet of the same length in this paper.
Based on a simplified handshaking process,
a CRA-based random access scheme was studied and analyzed
together with
a conventional CRA-based random access (which is a grant-free approach).
Since the length of payload 
can be adaptively decided depending on the number of active users,
it was shown that 
the CRA-based random access scheme 
with simplified handshaking process
can outperform others. We derived the throughput expressions
and showed that they agree with simulation results.
Thus, the throughput expressions
could be used 
to design a CRA-based random access system 
with IoT devices that have short 
messages to be delivered in MTC.

While we did not consider massive MIMO in this paper,
it is possible to generalize the approach with massive MIMO
based random access \cite{deC17}  \cite{Senel17} \cite{Liu18} \cite{Jiang15}
\cite{Ciuonzo15} in order to increase the number of users to be supported,
which might be a further research topic.

\appendices

\section{Proof of Lemma~\ref{L:1}}	\label{A:1}

From \eqref{EQ:UW}, it can be shown that
$B = \sum_{l=1}^L U_l + W_l$,
while $B_1 = \sum_{l=1}^L U_l$.
Then, 
under 
the assumption of {\bf A1},
$B$ and $B_1$ become binomial random variables as follows:
\begin{align}
\Pr(B_1 = b_1) & = \binom{L}{b_1} \alpha_1^{b_1} (1 - \alpha_1)^{L-b_1} \cr
\Pr(B = b) & = \binom{L}{b} (1 - \alpha_2)^b \alpha_2^{L-b}.
	\label{EQ:BB}
\end{align}
Since
$\uE[D_1 \,|\, B_1] = (1- P_{\rm MD}) B_1$,
from \eqref{EQ:BB}
we have
\be
\uE[D_1 \,|\, K] = (1- P_{\rm MD}) \uE[ B_1\,|\, K] =
(1- P_{\rm MD}) L \alpha_1.
	\label{EQ:ED1}
\ee
Using the same approach,
we have
\begin{align}
\uE[D_2 \,|\, K]
& = (1- P_{\rm MD}) \uE[ B_2\,|\, K] \cr
& = (1- P_{\rm MD}) \uE[ B - B_1\,|\, K] \cr
& = (1- P_{\rm MD}) L (1-\alpha_1 -\alpha_2).
	\label{EQ:ED2}
\end{align}
and
\begin{align}
\uE[D_3 \,|\, K] = P_{\rm FA} \uE[L-B \,|\, K] =
P_{\rm FA} L (1 - \alpha_2).
	\label{EQ:ED3}
\end{align}
From \eqref{EQ:ED1} --
\eqref{EQ:ED2}, we have \eqref{EQ:L1}, which completes the proof.

\section{Proof of Lemma~\ref{L:2}}	\label{A:2}

From \eqref{EQ:YT},
it follows that
\be
\bar Y = \tilde T_{\rm P} + T_{\rm D} \bar D,
	\label{EQ:YD}
\ee
where $\bar D = \uE[D] = \uE[\uE[D\,|\, K]]$.
Here, the first expectation is carried out over $K$
that has the distribution in \eqref{EQ:Kbeta_2}.
Using \eqref{EQ:Kbeta_2} and \eqref{EQ:L1},
after some manipulations,
it can be shown that
\be
\bar D = L \left((1-P_{\rm MD}) - e^{-\frac{\beta_2}{L}} (1 - P_{\rm MD}
- P_{\rm FA})\right).
	\label{EQ:bD}
\ee
In addition, from \eqref{EQ:YD}, we have
\be
\beta_2 = \lambda (\tilde T_{\rm P} + T_{\rm D} \bar D).
	\label{EQ:beta_2_1}
\ee
Substituting \eqref{EQ:bD} into \eqref{EQ:beta_2_1},
we have
\be
\frac{\beta_2}{L} = c_1 - c_2 e^{\frac{-\beta_2}{L}}.
	\label{EQ:bb}
\ee
Let $t = c_1 - \frac{\beta_2}{L}$. Then,
\eqref{EQ:bb} is re-written as
$t = c_2 e^t e^{-c_1}$ or
\be
-t e^{-t} = - c_2 e^{-c_1},
\ee
which implies that (by the definition of the Lambert W function)
\be
-t = \uW \left(- c_2 e^{-c_1}\right).
\ee
From this, we can obtain the expression
for $\beta_2$ in \eqref{EQ:L2}.
Substituting \eqref{EQ:bb} and \eqref{EQ:L2}
into \eqref{EQ:bD}, we have \eqref{EQ:L2b}.

\bibliographystyle{ieeetr}
\bibliography{mtc}

\end{document}